\newcommand{\BibitemShut}[1]{}

\documentclass[%
reprint,
 amsmath,amssymb,
 aps,
]{revtex4-1}

\usepackage{graphicx}
\usepackage{dcolumn}
\usepackage{bm}
\usepackage{natbib}

\begin{document}

\preprint{APS/123-QED}

\title{Production of rovibronic ground state RbCs molecules via two-photon cascade decay}
\author{Toshihiko Shimasaki} 
\author{Michael Bellos}   
\author{C. D. Bruzewicz}
\altaffiliation[Present Address: ]{Lincoln Laboratory, Massachusetts Institute of Technology, Lexington, Massachusetts 02420}
\author{Zack Lasner}
\author{David DeMille}
\affiliation{
Department of Physics, Yale University, New Haven, Connecticut 06520, USA\
}



%
\date{\today}
\begin{abstract}
We report the production of ultracold RbCs molecules in the rovibronic ground state, i.e., $X^1\Sigma^+ (v=0, J=0)$, by short-range photoassociation to the $2^3\Pi_{0}$ state followed by spontaneous emission. 
We use narrowband depletion spectroscopy to probe the distribution of rotational levels formed in the $X ^1\Sigma^+(v=0)$ state.  
We conclude, based on selection rules, that the primary decay route to $X ^1\Sigma^+(v=0)$ is a two-step cascade decay that leads to as much as $33\%$ branching into the $J=0$ rotational level.
The experimental simplicity of our scheme opens up the possibility of easier access to the study and manipulation of ultracold heteronuclear molecules in the rovibronic ground state.  
\end{abstract}

\pacs{Valid PACS appear here}
\maketitle


\section{\label{sec:level1}Introduction\protect\\} 

Tremendous efforts have been devoted to the production of ultracold samples of heteronuclear molecules over recent years. 
Promising applications range from quantum computation \cite{DeMilleQC} to ultracold chemistry \cite{JunYeUltracoldChemistry}, quantum dipolar physics \cite{Krems2009,JunYe2013}, and tests of fundamental physics \cite{Krems2009,DeMilleReview2009}.
However, access to dense samples of ground state ultracold polar molecules has so far been limited to a few outstanding experiments \cite{JunYeSTIRAP,InnsbruckRbCs2014}, which utilized magneto-association followed by transfer to the ground state by stimulated Raman adiabatic passage \cite{RevModPhys.70.1003}.
A less experimentally complex path toward a sample of ultracold molecules is short-range photoassociation (PA) \cite{Weidemuller2008}.
Recent discoveries of short-range PA transitions in RbCs \cite{Gabbanini2011}, NaCs \cite{PhysRevA.84.061401}, KRb \cite{PhysRevA.86.053428} and Rb$_2$ \cite{PhysRevA.87.053404,Bellos2011} proved the general applicability of the process to alkali dimers. 
Short-range PA is a convenient means to produce ultracold molecules due to its simplicity and possible continuous operation. Although PA leads to an unavoidable distribution of molecules over many vibrational, rotational, and hyperfine levels, it has been argued that simple measures may allow removal of excited states following PA \cite{ERHudson2008,Bruzewicz14}.
However, to date short-range PA has predominantly produced molecules in rotationally excited states: $\sim 2$\% of molecules were observed in $J=0$ for LiCs \cite{Weidemuller2008} and no $J=0$ molecules were observed for NaCs \cite{PhysRevA.84.061401}.

\begin{figure*}[htp]
\includegraphics[scale=0.32]{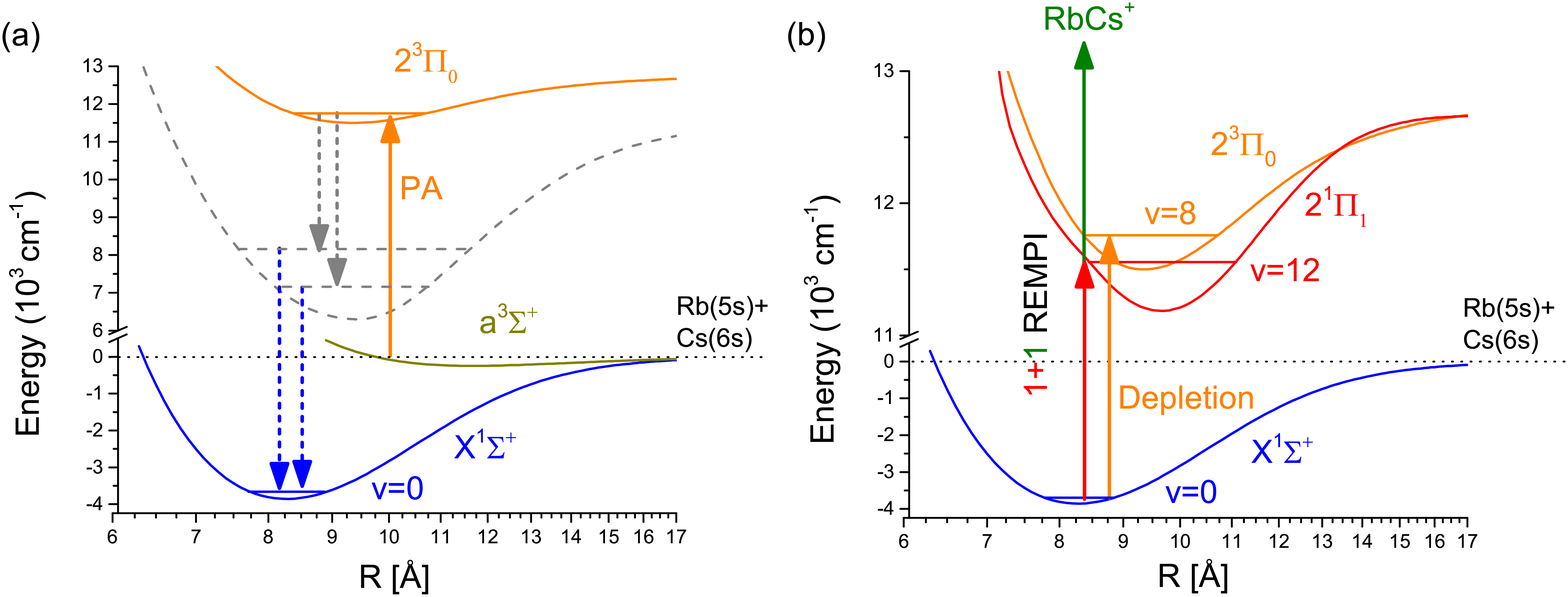}
\caption{\label{fig:formationdetectionpathway} (Color online) (a) Short-range photoassociation followed by two-photon cascade decay to rovibronic ground state molecules. 
(b) Detection of vibronic ground state molecules through two-color REMPI. Also shown is the depletion transition used for rotational spectroscopy. Here, a narrow-linewidth laser depopulates given rotational levels from $X(v=0)$ and thus modulates the REMPI signal. Potential energy curves are from Ref. \cite{Allouche2002}.
}
\end{figure*}

In this paper, we perform high-resolution depletion spectroscopy \cite{WangPRA2007,Weidemuller2008} to measure the distribution of rotational levels in RbCs molecules produced via short-range PA to the $2^3\Pi_0$ electronic state. 
We confirm that a large fraction of these $X(v=0)$ molecules, up to 33 \%, are in their rovibronic ground states, i.e., $X$($v=0$, $J=0$).
We also show that the formation pathway to $X(v=0)$ is a two-photon cascade decay as shown in Fig. \ref{fig:formationdetectionpathway}(a), 
as opposed to a direct one-photon decay as was previously supposed \cite{Bruzewicz14}.
\section{\label{sec:level1}Experimental setup\protect\\}
Most of our experimental setup has been previously described \cite{Bruzewicz14}.
$^{85}$Rb and Cs atoms are laser-cooled and trapped in a dual-species forced dark spontaneous force optical trap (dark SPOT) \cite{KetterleDarkSPOT} loaded by alkali dispensers.
The overlap of the two atom clouds is optimized using absorption imaging from two orthogonal directions.
The typical density $n$ and total atom number $N$ for Rb and Cs are $n_{{\rm Rb}}\sim 6 \times 10^{10}$ cm$^{-3}$, $N_{{\rm Rb}} \sim 5 \times 10^6$, $n_{{\rm Cs}} \sim 8 \times 10^{10}$ cm$^{-3}$, and $N_{{\rm Cs}} \sim 1 \times 10^7$. 
The translational temperature of the atoms is measured by time-of-flight imaging to be $T \sim 100\ \mu$K.
For the PA transition, we use up to 300 mW of light from a Ti:sapphire laser, focused onto the atom clouds with a beam waist ($1/e^2$ power radius) of 100 $\mu$m. 
We use four distinct PA transitions in this work, namely to the first two rotational levels of the $2^3\Pi_{0^+}(v=10)$ state ($J^P=0^+$ and $1^-$ near 11817.1 cm$^{-1}$) \cite{Bruzewicz14} and of the $2^3\Pi_{0^-}(v=11)$ state ($J^P=0^-$ and $1^+$ near 11803.9 cm$^{-1}$) \cite{Gabbanini2013}. Here $J$ corresponds to the total angular momentum excluding nuclear spin, and $P$ to the parity.

We detect molecules using two-color resonance-enhanced multi-photon ionization (1+1 REMPI) as shown in Fig. \ref{fig:formationdetectionpathway}(b). 
The first photon from a nanosecond pulse dye laser resonantly excites the vibronic ground state to a given excited state ($X(v=0)$ $\rightarrow $ 2$^1\Pi_1 (v=12)$) at 15342.0 cm$^{-1}$ \cite{Gustavsson1988}, 
where an intense ($\sim 2$ mJ/pulse) 532 nm pulse ionizes the molecules 10 ns later. 
The energy of the resonant pulse is kept below 0.1 mJ to minimize power broadening and off-resonance excitation. 
The molecular ions are accelerated by the electric field inside the vacuum chamber ($\sim 100$ V/cm) toward a Channeltron detector, where they are converted into electrical signals. The RbCs$^+$ signals are separated from other atomic and molecular signals based on time-of-flight measurements.
The experiment is repeated at 100 Hz, the repetition rate of the REMPI lasers.
Since the linewidth of the dye laser ($\sim 6$ GHz) is larger than the spacing of rotational levels, it is impossible to address molecules in a single rotational level.

\begin{figure}
\includegraphics[scale=0.4]{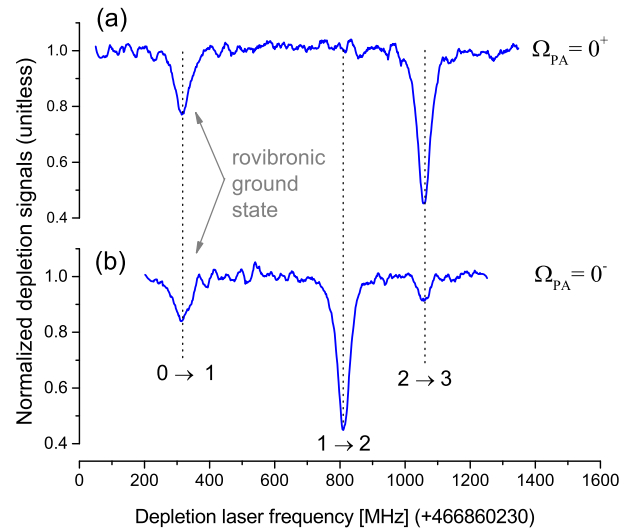}
\caption{\label{fig:depletionspectra} (Color online) Depletion spectra with the PA laser tuned to (a) the $|\Omega_{{\rm PA}}=0^+,\ J_{{\rm PA}}^P=0^+ \rangle$ line and (b) the $|\Omega_{{\rm PA}}=0^-,\ J_{{\rm PA}}^P=0^- \rangle$ line.
The normalized depletion signal is taken at 2 MHz frequency steps with an average over 400 pulses and shown after smoothing \cite{Fig2comment}.
Labels indicate the rotational quantum numbers $J_X \rightarrow J'$ of the depletion transition. The sum of depletion depths is over 90\%, 
which indicates that the ion signal originating from off-resonant excitation or from excited vibrational levels is small in our REMPI spectroscopy. Shown here are several close-lying $J \rightarrow J' = J+1$ lines; we also observed all associated $J \rightarrow  J' = J-1$ lines, at their predicted frequencies.}
\end{figure}

To circumvent this difficulty, we use the technique of high-resolution depletion spectroscopy \cite{WangPRA2007, Weidemuller2008}. 
An additional CW diode laser is used to deplete the population of any given rotational level in the $X(v=0)$ manifold of states. 
The $2 ^3\Pi_{0^+}(v=8)$ level is chosen as the upper state of this depletion transition, 
as all molecular constants for the upper and lower states of this transition are accurately known \cite{JTKim2008,Fellows1999}, making rotational line assignments unambiguous.
Following the excitation, molecules decay predominantly into other vibrational states before the REMPI detection pulse arrives. The molecular ion signal is monitored while the frequency of the depletion laser is scanned; the population in a particular rotational level appears as a decrease in the molecular ion signal as shown in Fig. \ref{fig:depletionspectra}. 
 
To reduce the effect of drifts in the ion signal, we cycle the depletion laser on and off for consecutive detection pulses. We separately average signals with and without the depletion laser after 400 cycles 
and take a ratio of these two averages to obtain a normalized depletion signal (Fig. \ref{fig:depletionspectra}).
\section{\label{sec:level1}Results\protect\\}
Initially, we tuned the PA laser to the $2 ^3\Pi_{0^+}(v=10)$ state, which we found in \cite{Bruzewicz14} to be one of the most efficient states for producing $X(v=0)$ state molecules.
For PA to the $J^P_{{\rm PA}}=0^+$ level, one-photon decay would lead to the production of exclusively $J^P_X=1^-$ molecules due to the  $\Delta J \leq 1,\ J=0 \not\leftrightarrow 0$, and $P_iP_f=-1$ selection rules [where $P_i$ $(P_f)$ is the parity of the initial (final) state of the transition].
However, in a scan of the depletion laser, we observed two clear peaks, which we identified as originating from $J^P_X=0^+$ and $J^P_X=2^+$ molecules in the ground state (Fig. \ref{fig:depletionspectra}(a)).
The absolute positions of, and rotational splittings between, all observed lines agree (within our experimental uncertainties of $\sim$300 MHz and $\sim$30 MHz respectively) with the values predicted from the molecular constants of Ref. \cite{JTKim2008,Fellows1999}.
We obtained a similar result with the PA laser tuned to the $J^P_{{\rm PA}}=1^-$ line; 
here we observed two peaks originating from $J^P_X=1^-$ and $J^P_X=3^-$.
These observations suggest that two-photon decays are solely responsible for the population of the $X(v=0)$ state.

Then, we tuned the PA laser to the $2 ^3\Pi_{0^-}$ state observed by Fioretti and Gabbanini \cite{Gabbanini2013}.
As is discussed below, our previous assignment \cite{Bruzewicz14} of this PA series to an $\Omega=0^-$ potential is consistent with our results presented here.
For the  $J^P_{{\rm PA}}=0^-$ line, we mainly observed $J^P_X = 1^-$ molecules, 
which would be the only state populated if the parity changes in each step of a two-photon cascade.
However, we also observed small peaks corresponding to $J^P_X=0^+$ and $J^P_X=2^+$ as shown in Fig. \ref{fig:depletionspectra}(b).
Similarly, for the $J^P_{{\rm PA}} =1^+$ line, we observed population in all of the rotational levels possible by a two-photon decay but without obeying the usual parity selection rule: $J^P_X=0^+,1^-,2^+$, and $3^-$.
\begin{figure*}[htp]
\includegraphics[scale=0.32]{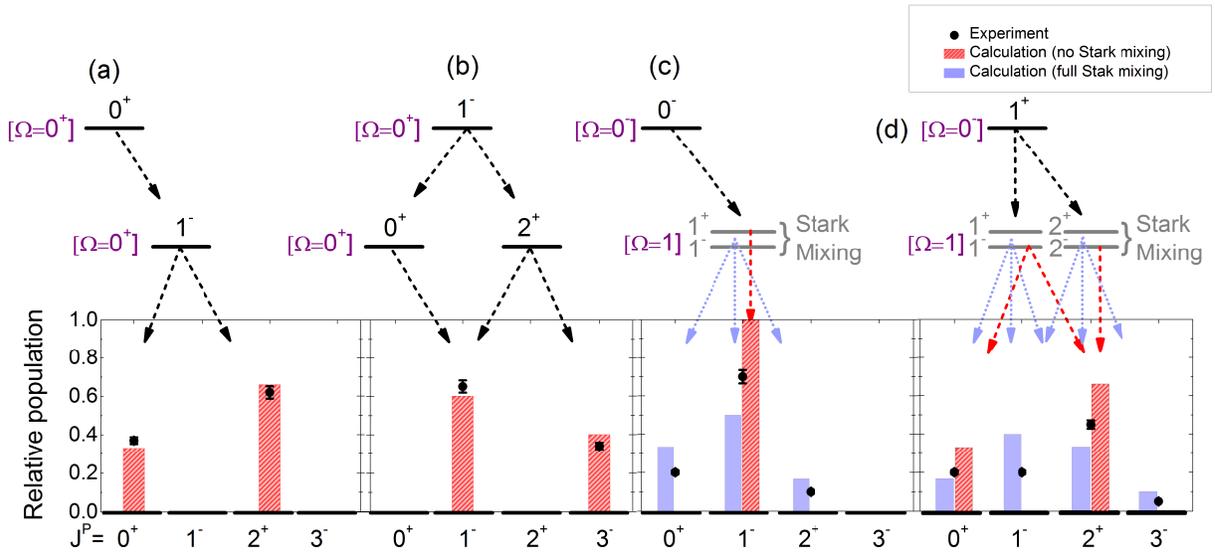}
\caption{\label{fig:Rotationaldistribution} (Color online) Decay pathways and ground state rotational distribution starting from four distinct PA states.
Molecular states are labeled by their $\Omega$ and $J^P$ quantum numbers.  
Black squares are the measured population distribution and colored histograms show calculated distributions  
based on H${\rm \ddot{o}}$nl-London factors. 
Shaded bars are obtained assuming parity selection rules are followed throughout the decay, i.e. no Stark mixing in intermediate levels, and match the results for the $\Omega=0^+$ PA state.
For the $\Omega_{{\rm PA}} =0^-$ PA state, solid bars are calculated assuming complete Stark mixing of $\Omega$-doublets in the intermediate state.
The measured values fall between the two predictions for the extreme assumptions; hence Stark mixing of the intermediate $\Omega$-doublet states is substantial, but not complete.
 }
\end{figure*}

The qualitative difference in the decay of the $\Omega =0^+$ and $\Omega =0^-$ PA states arises from the different intermediate states of the cascade.
Due to the $\Omega=0^+ \not\leftrightarrow \Omega=0^-$ selection rule,
$\Omega=0^-$ states cannot decay directly to $\Omega = 0^+$ states, including the ground $X^1\Sigma^+$ state. 
Thus, the only possible decay path for $\Omega = 0^-$ molecules to the $X^1\Sigma^+$ state is a two-step decay via intermediate states with $\Omega =1$.
These intermediate $\Omega = 1$ states have $\Omega$-doublet substructure, such that for each $J$ there is a pair of nearly-degenerate levels of opposite parity.  In our experiment, these levels are likely to be mixed due to the electric field used to extract ions for detection.  Hence, for decays via such $\Omega =1$ intermediate states, the usual parity selection rules are no longer valid under our experimental conditions.
By contrast, the $\Omega = 0^+$ excited state can decay to the $X$ state via a cascade with intermediate states that have either $\Omega = 0^+$ or $\Omega = 1$.  Our observations indicate that the overwhelmingly dominant pathway is through an $\Omega = 0^+$ intermediate state. 
We summarize the possible decay paths in Fig. \ref{fig:Rotationaldistribution},
along with the distribution of the rotational levels obtained by analyzing depletion spectra.
We renormalize the sum of the experimental population fractions to 1 to simplify the comparison with theoretical predictions.
The observed rotational distributions are consistent with predictions from H$\ddot{{\rm o}}$nl-London factors \cite{HonlLondon1974} for decays of the $\Omega = 0^+$ PA state, assuming that each molecule decays only  through an $\Omega = 0^+$ intermediate state. However, for $\Omega=0^-$ PA states, the unknown degree of parity mixing in the intermediate $\Omega=1$ states due to the Stark effect makes a quantitative prediction of rotational distributions difficult.
Instead, we calculate rotational distributions for two limiting cases: (1) assuming that the parity selection rule is exactly satisfied throughout the decay (the case of no Stark mixing) and (2) assuming that the two intermediate states are fully mixed, resulting in maximal population of otherwise parity-forbidden levels. The experimental values for the rotational distributions fall between these two limiting cases as shown in Fig. \ref{fig:Rotationaldistribution} (c) and (d).

Similar two-photon cascade decays have been observed in Cs$_2$ \cite{PhysRevA.79.021402,PhysRevA.85.030502}, where some of the one-photon direct decays are strictly forbidden by the ungerade-gerade selection rule applicable to homonuclear diatomic molecules. 
In RbCs, the intercombination transitions from the $2 ^3\Pi_{0^+}$ state (a triplet state) to the $X$ state (a singlet state) should be allowed due to singlet-triplet mixing by the spin-orbit interaction.
However, we did not observe any one-photon decay to $J^P_X=1^-$ molecules for the $|\Omega_{{\rm PA}}=0^+, 
J^P_{{\rm PA}}=0^+ \rangle$ line (Fig. \ref{fig:Rotationaldistribution}(a)), which indicates that the one-photon decay path is much less probable than the two-photon cascade.
We suspect that the two-photon cascade for $\Omega_{{\rm PA}} = 0^+$ occurs via states in the $A^1\Sigma^+[0+] / b^3\Pi[0+]$ complex,
where singlet-triplet mixing could be very efficient due to the avoided crossing between these two potential curves \cite{PhysRevA.81.042511}.
However, we do not have a direct way to confirm this hypothesis.

\begin{figure}
\includegraphics[scale=0.4]{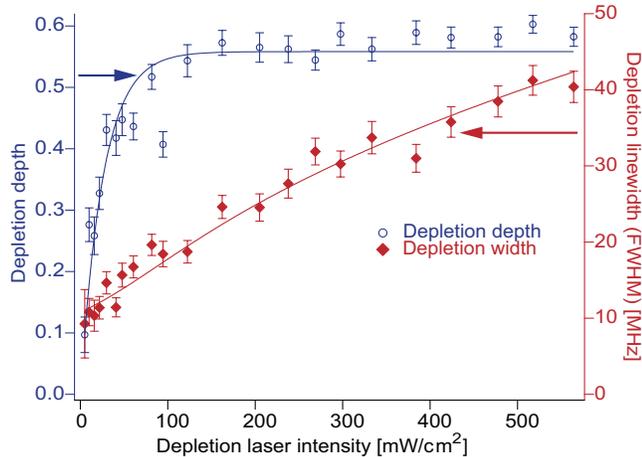}
\caption{\label{fig:SaturationBehavior} (Color online) Power dependence of the depletion signal depth and linewidth of the depletion feature for the $J_X=2 \rightarrow J' = 3$ transition. 
Data points and error bars are extracted from curve fits of corresponding depletion scans. 
Lines are fits to Eqn.(1) with $\langle \mu \rangle$ as a free parameter. 
The best fit values are $ \langle \mu \rangle = 1.7 \times 10^{-2} \ e a_0 \ ( 1.5 \times 10^{-2} \  e a_0)$ for the depletion depth (depletion linewidth) data. The deviation of the data points of the depletion depth from the fit curve at high depletion laser intensities likely arises from the imperfect assumption of a single interaction time.}
\end{figure}

To independently verify that the one-photon spontaneous emission probability is small, we determine the transition dipole matrix element $\langle \mu \rangle$ between the $X$ and $2 ^3 \Pi_{0^+}$ states by measuring the saturation behavior of the depletion transition as shown in Fig. \ref{fig:SaturationBehavior}.
This behavior can be understood by modeling the depletion transition as an open two-level system. 
Since the decay rate from the upper state is much higher than the excitation rate by the depletion laser, we can ignore coherent effects such as Rabi oscillations. We also assume a single average interaction time $\tau \sim 10$ ms, consistent with the transit time through the depletion laser beam determined by thermal velocities or by free fall in gravity.
In this model, the depletion lineshape $L$ is given by
\begin{equation}
L(\Delta, \Omega_R ) =A \left\{ 1- \exp \left( -\frac{(\frac{\gamma}{2})^2}{\Delta^2 + (\frac{\gamma}{2})^2}   \frac{\Omega_{R}^2}{\gamma} \tau \right) \right\},
\end{equation}
where $\Delta$ is the detuning of the depletion laser from resonance,  $\Omega_{R} = E \cdot \langle \mu \rangle \chi / \hbar$ is the Rabi frequency for a transition dipole moment $\langle \mu \rangle$ driven by the electric field amplitude $E$ of the depletion laser, $\chi \approx 0.12$ is the factor arising from the nuclear wavefunctions (both the Franck-Condon factor and the angular factor), 
and $\gamma$ is the natural linewidth of the transition.
   
With this model, we analyzed the power dependence of the depletion depth and the linewidth of the transition as shown in Fig. \ref{fig:SaturationBehavior}. By fitting the data with Eq. (1), we obtain $\langle \mu \rangle \approx 2 \times 10^{-2} \ e a_0$.
We estimate this value to have an overall uncertainty of a factor of $\sim 2$, dominated by the uncertainty in our estimate of the interaction time.
In Ref. \cite{Bruzewicz14}, a crude argument led to an estimated value of this matrix element of $0.3 \ ea_0$. 
Our result suggests that the one-photon direct decay is less efficient than we previously expected in \cite{Bruzewicz14}.
This is consistent with our non-observation of any single-photon decay.

Finally, in \cite{Bruzewicz14}, we estimated the production rate of $X(v=0)$ molecules to be $\sim 6\times 10^3$ molecules/s starting from atoms in a dark SPOT. By considering the branching ratio (1/3) for the newly identified pathway, we conclude that the production rate of $X(v=0,J=0)$ molecules via the $|\Omega=0^+, J^P_{{\rm PA}} = 0 \rangle$ PA line is $\sim 2 \times 10^3$ molecules/s. This is the highest production rate of molecules in the rovibronic ground state via PA to date.
\section{\label{sec:level1}Conclusion\protect\\}
We performed depletion spectroscopy to resolve the rotational level distribution of RbCs molecules produced in the $X(v=0)$ state following short-range PA to the $2 ^3 \Pi_{0^+}$ and $2 ^3 \Pi_{0^-}$ states. 
In contrast to our expectations (as discussed in Ref. \cite{Bruzewicz14}), we found that the $X$ state population arises from a two-photon cascade decay in both cases, rather than a direct one-photon decay from the $2^3\Pi_{0^+}$ state. 
We also confirmed that PA to the $2 ^3 \Pi_{0^+}, J^P_{{\rm PA}}=0$ lines produce a substantial fraction ($\approx$33\%) of $X(v=0)$ state molecules in the rotational ground state at a rate of $\sim 2 \times 10^3$ molecules/s. 
This represents the strongest demonstrated pathway to the production of rovibronic ground state molecules via PA to date.
This demonstration is a starting point for accumulating ultracold molecules in the rovibronic ground state by PA in an optical trap.
There, the higher densities and lower temperatures achievable should allow substantially higher production rates; moreover, collisions with co-trapped atoms should remove metastable excited states.  Hence our results point to the possibility of a simple procedure for producing a pure trapped sample of rovibronic ground state molecules.
\begin{acknowledgments}
This work was supported by DOE, AFSOR MURI, and ARO MURI.
\end{acknowledgments} 
\bibliography{DepletionPaper}

\begin{thebibliography}{28}
\bibitem [{Fig()}]{Fig2comment}%
  \BibitemOpen
  \href@noop {} {\ }\bibinfo {note} {The data points are smoothed by a
  Savitzky-Golay filter with a window size of 20 points and cubic
  polynomials.}\BibitemShut {Stop}%
\end{thebibliography}
\end{document}